\documentclass[journal, twoside]{IEEEtran}
\usepackage[T1]{fontenc}
\usepackage{lineno}
\usepackage{tikz}
\usepackage{pgfplots}
\usepackage{pgfplotstable}
\usepackage{amsmath,amssymb}
\usepackage[ backend=biber,style=ieee,sorting=none, maxnames = 10]{biblatex}

\renewcommand\footnotemark{}

\bibliography{sample}

\DeclareBibliographyCategory{ignore}

\makeatletter
\def\blfootnote{\gdef\@thefnmark{}\@footnotetext}
\makeatother

\definecolor{C1}{RGB}{031, 119, 180} 
\definecolor{C2}{RGB}{255, 127, 014} 
\definecolor{C3}{RGB}{044, 160, 044} 
\definecolor{C4}{RGB}{215, 039, 040} 
\definecolor{C6}{RGB}{148, 103, 189} 
\definecolor{C100}{RGB}{140, 086, 075} 
\definecolor{C7}{RGB}{227, 119, 194} 
\definecolor{C8}{RGB}{127, 127, 127} 
\definecolor{C9}{RGB}{188, 189, 034} 
\definecolor{C5}{RGB}{023, 190, 207} 

\definecolor{C11}{RGB}{174, 199, 232} 
\definecolor{C12}{RGB}{255, 187, 120} 
\definecolor{C13}{RGB}{152, 223, 138} 
\definecolor{C14}{RGB}{255, 152, 150} 
\definecolor{C15}{RGB}{197, 176, 213} 
\definecolor{C16}{RGB}{196, 156, 148} 
\definecolor{C17}{RGB}{247, 182, 210} 
\definecolor{C18}{RGB}{199, 199, 199} 
\definecolor{C19}{RGB}{219, 219, 141} 
\definecolor{C20}{RGB}{158, 218, 229} 

\tikzstyle{gradientc1} = [bottom color=C1!5, top color=C1!15]
\tikzstyle{gradientc4} = [bottom color=C4!5, top color=C4!15]    
\tikzstyle{gradientc6} = [bottom color=C6!5, top color=C6!15]
\tikzstyle{gradientc8} = [bottom color=C8!5, top color=C8!15]

\pgfplotscreateplotcyclelist{foo}{
        {thick,C1,mark=*},
        {thick,C2,mark=square*},
        {thick,C3, mark = triangle*},
        {thick,C4,mark=+},
        {thick,C5, mark = diamond*},
        {thick,C6, mark = x},
        {thick,C7, mark = pentagon*},
        {thick, C8, mark = 10-pointed star},
        {thick, C9, mark = Mercedes star},
        {thick, C10, mark = heart}
      }

\pgfplotscreateplotcyclelist{foo2}{
        {thick,C1,mark=*},
        {thick,C2,mark=square*},
        {thick,C3, mark = triangle*,dotted},
        {thick,C4,mark=+,dashed},
        {thick,C5, mark = diamond*,dashed},
      }
\begin{document}

\title{Information Reconciliation for Continuous-Variable Quantum Key Distribution with $\beta > 1$ Using Short Blocklength Error Correction Codes: \newline Proposal and Concerns
}

\author{Kadir G\" um\" u\c s, João dos Reis Frazão, Aaron Albores-Mejia, Boris \v{S}kori\'{c}, Gabriele Liga, \\ Yunus Can G\"{u}ltekin, Thomas Bradley, and Chigo Okonkwo}

 \thanks{This work was supported by the Dutch Ministry of Economic Affairs and Climate Policy (EZK), as part of the PhotonDelta National GrowthFunds Programme on Photonics and the QuantumDeltaNL National Growthfunds on Quantum Technology. \emph{(Corresponding author: Kadir G\" um\" u\c s})}
 \thanks{Kadir G\" um\" u\c s, João dos Reis Frazão,  Aaron Albores-Mejia, Thomas Bradley, and Chigo Okonkwo are with the High Capacity Optical Transmission Laboratory, Electro-Optical Communications Group, Eindhoven University of Technology, 5600 MB, Eindhoven, The Netherlands. (e-mails: k.gumus@tue.nl, j.c.dos.reis.frazao@tue.nl, a.albores.mejia@tue.nl, t.d.bradley@tue.nl, cokonkwo@tue.nl)}
\thanks{Aaron Albores-Mejia and Chigo Okonkwo are with CubiQ Technologies, De Groene Loper 5, Eindhoven, The Netherlands (e-mails: aaron@cubiq-technologies.com, chigo@cubiq-technologies.com)}
\thanks{Boris \v{S}kori\'{c} is with the Department of Mathematics and Computer Science, Eindhoven University of Technology, 5600 MB, Eindhoven, The Netherlands (e-mail: b.skoric@tue.nl)}
 \thanks{Gabriele Liga and Yunus Can G\" ultekin are with the Information and Communication Theory Lab, Signal Processing Group, Eindhoven University of Technology, 5600MB, The Netherlands (e-mails: g.liga@tue.nl, y.c.g.gultekin@tue.nl)}

\maketitle

\begin{abstract} 
In this paper, we introduce a reconciliation protocol with a two-step error correction scheme that uses a short-blocklength, low-rate code and a long-blocklength, high-rate code. We simulate the protocol using a short-block-length low-density parity-check code and show that we can achieve reconciliation efficiencies above 1 using this method. We discuss the necessary steps required regarding security proofs before this protocol can be securely implemented.
\end{abstract}

\section{Introduction}
Concerns about data security have been growing over the past couple of years with the advent of quantum computing \cite{gyongyosi2019survey}, and as a result, quantum key distribution (QKD), first proposed in \cite{Bennett_2014}, has become a widely researched topic. Powerful enough quantum computers could break existing cryptography protocols using Shor's algorithm \cite{shor1999polynomial}.  
QKD enables the exchange of unconditionally secure keys between two communicating parties, Alice and Bob. If an eavesdropper Eve attempts to eavesdrop on the quantum channel, Alice and Bob will notice and abort the key exchange. 

In general, QKD is categorised into two different streams: discrete-variable (DV) \cite{Bennett_2014}, and continuous-variable (CV) QKD \cite{GG02}. The main difference lies in the measurement of quantum states: in DV-QKD, single photons are measured, whereas in CV-QKD, a significantly attenuated coherent signal is detected. The advantage of CV-QKD is that standard telecommunication components can be used for implementation, allowing for a more cost-effective product that is easier to integrate into the current telecommunication network. On the other hand, for DV-QKD, expensive single-photon detectors are required \cite{Laudenbach_2018}. Where DV outshines CV, however, is in the complexity of the post-processing. For DV-QKD, post-processing is relatively simple, whereas for CV-QKD, it is a major system bottleneck \cite{yang2023information}. 

An essential part of the post-processing for CV-QKD is the reconciliation. The goal of reconciliation is to perform error correction to allow for the exchange of bits between Alice and Bob using the transmitted and measured quantum states in a secure manner. These bits will be used to distil the key during privacy amplification. Multi-dimensional reconciliation, introduced in \cite{Leverrier_2008}, is a popular choice for long-distance links, whereas for shorter-distance links, slice reconciliation \cite{Assche2004} is preferred. Other reconciliation protocols have been proposed as well, such as a rate-adaptive protocol \cite{wang2017efficient}, one involving multiple decoding attempts \cite{gumucs2021novel}, and a protocol using random codebooks \cite{ray2025random}. 

The performance of these error-correction codes used during reconciliation determines both the achievable secret key rates (SKRs) and the distance of the CV-QKD protocol. Therefore, the error-correcting codes used have long blocklengths, operating close to the Shannon capacity \cite{yang2023information}. Additionally, due to the low signal-to-noise ratio (SNR) of the quantum channel, the error-correction code rates are low, making their decoding complex \cite{Milicevic_2018}.
In \cite{mani2021multiedge, gumucs2021low}, low-rate low-density parity-check (LDPC) codes were designed for reconciliation. Raptor codes have been studied in \cite{Zhou2019}, Polar codes have been studied in \cite{Jouguet2012}, and recently LDPC codes concatenated with Polar codes have been proposed in \cite{cao2023ic}. 
Because the decoding of these codes is complex, the information throughput is significantly lower than in the rest of the CV-QKD system, thereby limiting the practically achievable key rates.

The reconciliation efficiency, $\beta$, plays a significant role in the performance of a CV-QKD system. 
The reconciliation efficiency is a measure of how close the error-correction performance is to the Shannon capacity and is defined as the code rate $R$ divided by the quantum channel capacity $I_{AB}$. Normally, the assumption is that $\beta$ is bounded by 1, as it is impossible to reliably transmit information at a higher rate than the Shannon capacity \cite{shannon1948mathematical}. During reconciliation, the frame error rate (FER) of the error correction is the fraction of frames which are rejected. In many works \cite{Milicevic_2018,gumucs2023adaptive,Jouguet2012}, this FER is allowed to be arbitrarily high, as we can simply discard any incorrectly decoded frame. On the other hand, the Shannon limit holds when an arbitrarily low FER is desired. Therefore, if high FER is allowed, it is possible to transmit data using error correction codes with $\beta > 1$ without violating the Shannon limit. The total amount of reliably transmitted data is below the Shannon limit because of the high FER.

In this work, we propose a new reconciliation protocol that involves a two-stage decoding process, combining a low-rate, short-blocklength error correction code and a high-rate, long-blocklength error correction code. We show that by rejecting most frames, it is possible to achieve reconciliation efficiencies above 1. We also discuss the potential security concerns with the protocol that need to be resolved before any conclusions can be drawn about the secret key rate or the achievable distance. 

The remainder of the paper is organised as follows. We describe our proposed protocol in Section~\ref{Section:Proposed Reconciliation}, while we simulate short blocklength LDPC codes to validate our protocol in Section \ref{Section: Results}. We discuss the implications on the security and the secret key rate in Section \ref{Section:Security concerns}. 
Finally, we conclude our paper in Section~\ref{Section:Conclusion} and propose further research avenues. 

\section{Multi-dimensional Reconciliation}
\begin{figure}[!t]
    \centering
    \resizebox{\linewidth}{!}{\tikzstyle{gradientc1} = [bottom color=C1!5, top color=C1!15]
\tikzstyle{gradientc4} = [bottom color=C4!5, top color=C4!15]    
\tikzstyle{gradientc6} = [bottom color=C6!5, top color=C6!15]
\tikzstyle{gradientc8} = [bottom color=C8!5, top color=C8!15]
\tikzstyle{gradientc7} = [bottom color=C7!5, top color=C7!15]
\begin{tikzpicture}[>=latex]

\draw[rounded corners,dashed, C6](-2,2.35) rectangle (0.15,5.65);

\begin{scope}[shift = {(-0.5,0)}]
\draw[C1,->,thick](-0.5,2) -- node[left, yshift = -10.75mm, xshift = -1mm]{$\mathbf{x}$}(-0.5,6);
\draw[thick,C6,gradientc6] (-0.5,4) circle (0.3cm) node {+};
\draw[thick,C6,->](-1.45,4) -- node[midway,above, yshift  = 1mm]{$\mathbf{z}$}(-0.8,4);
\end{scope}
\node[rotate = 90,C6] at (-0.15,4){Quantum Channel};

\draw[rounded corners, C8, dashed] (0.25,2.35) rectangle(11.25,5.65) {};
\node[C8] at (5.75,4) {Classical Channel};

\draw[very thick, rounded corners, color = C1,dashed] (-2,0.25) rectangle (11.25,2.25);
\node at (4.625,0.5){\textcolor{C1}{Alice}};

\begin{scope}[shift = {(-1,0)}]
\draw[rounded corners,C1,gradientc1,thick] (-0.75,1) rectangle node[midway, align = center, font=\footnotesize]{Coherent \\ Quantum \\ Transmitter} (0.75,2); 

\draw[rounded corners, draw = C1, line width = 0.4mm,gradientc1] (1.5,1) rectangle (3,2);

\draw[rounded corners, draw = C1, line width = 0.4mm,gradientc1] (3.75,1) rectangle (5.25,2);

\draw[rounded corners, draw = C1, line width = 0.4mm,gradientc1] (6,1) rectangle (7.5,2);

\draw[rounded corners, draw = C1, line width = 0.4mm,gradientc1] (10.5,1) rectangle (12,2);

\node at (6.75,1.5){\footnotesize \textcolor{C1}{Decode}};
\node at (4.5,1.5){\footnotesize \textcolor{C1}{LLR}};
\node at (2.25,1.5){\footnotesize \textcolor{C1}{Demap}};
\node at (11.25,1.5){\footnotesize \textcolor{C1}{Hashing}};

\draw[->,thick,draw = C1] (0.75,1.5) -- (1.5,1.5);
\node at (1.125,1.75) {\textcolor{C1}{$\mathbf{x}$}};

\draw[->,thick,draw  =C1] (3,1.5) -- (3.75,1.5);
\node at (3.375,1.75){\textcolor{C1}{$\mathbf{r}$}};

\draw[->,thick, draw = C1] (5.25,1.5) -- (6,1.5);
\node at (5.625,1.75){\textcolor{C1}{$\mathbf{l}$}};

\draw[->,thick,draw = C1] (7.5,1.5) -- (10.5,1.5);
\node at (9,1.75) {\textcolor{C1}{$\mathbf{\hat{s}}$}};

\end{scope}
\draw[very thick, rounded corners, color = C4,dashed] (-2,7.75) rectangle (11.25,5.75);
\node at (4.625,7.5){\textcolor{C4}{Bob}};

\begin{scope}[shift = {(-1,0)}]

\draw[rounded corners,C4,gradientc4,thick] (-0.75,6) rectangle node[midway, align = center,font=\footnotesize]{Coherent \\ Quantum \\ Receiver} (0.75,7); 

\draw[rounded corners,  draw = C4, line width = 0.4mm,gradientc4] (1.5,7) rectangle (3,6);

\draw[rounded corners, draw = C4,line width = 0.4mm,gradientc4] (6,6) rectangle (7.5,7);

\draw[rounded corners, draw = C4,line width = 0.4mm,gradientc4] (3.75,6) rectangle (5.25,7);

\draw[rounded corners, draw = C4, line width = 0.4mm,gradientc4] (8.25,6) rectangle (9.75,7);

\draw[rounded corners, draw = C4, line width = 0.4mm,gradientc4] (10.5,6) rectangle (12,7);

\node at (9,6.5){\footnotesize \textcolor{C4}{QRNG}};
\node at (6.75,6.5){\footnotesize \textcolor{C4}{Encode}};
\node at (4.5,6.5){\footnotesize \textcolor{C4}{BPSK}};
\node at (2.25,6.5){\footnotesize \textcolor{C4}{Map}};
\node at (11.25,6.5){\footnotesize \textcolor{C4}{Hashing}};

\draw[->,thick, draw  =C4] (0.75,6.5) -- (1.5,6.5);
\node at (1.125,6.25) {\textcolor{C4}{$\mathbf{y}$}};

\draw[->,thick, draw = C4] (8.25,6.5) -- node[midway,below, yshift =-0.25mm]{\textcolor{C4}{$\mathbf{s}$}} (7.5,6.5);

\draw[->,thick, draw = C4] (9.75,6.5) -- node[midway,below, yshift =-0.25mm]{\textcolor{C4}{$\mathbf{s}$}} (10.5,6.5);

\draw[->,thick,draw = C4](6,6.5) -- node[midway,below,C4,yshift = -0.25mm]{$\mathbf{c}$}(5.25,6.5);

\draw[<-,thick,draw = C4] (3,6.5) -- (3.75,6.5);
\node at (3.375,6.25){\textcolor{C4}{$\mathbf{u}$}};

\end{scope}
\draw[->,thick, draw = C4](1.25,6) -- (1.25,2);
\node at (1,4){\textcolor{C4}{$\mathbf{m}$}};

\draw[->,thick, draw = C4](10.5,6) -- (10.5,2);
\draw[->,thick, draw = C1](10,2) -- (10,6);
\node at (9.65,4){\textcolor{C1}{$h_{\mathbf{\hat{s}}}$}};
\node at (10.85,4){\textcolor{C4}{y/n}};

\end{tikzpicture}}
    \caption{An overview of multi-dimensional reconciliation.}
    \label{Reconciliation}
\end{figure}
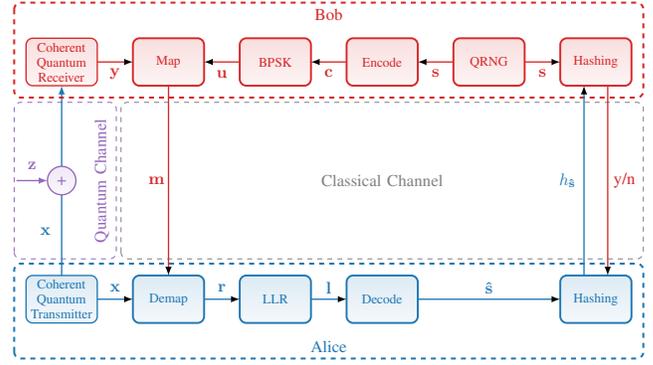

Multi-dimensional reconciliation, first introduced in \cite{Leverrier_2008}, is a commonly used reconciliation protocol, especially for long-distance CV-QKD systems \cite{yang2023information}.
As direct reconciliation is limited by the 3~dB limit \cite{Laudenbach_2018}, we will only consider reverse reconciliation.
The goal of the reconciliation is to share a string of bits $\mathbf{s}$, which will be used to distil the keys in the privacy amplification, between Alice and Bob, such that they have more information on $\mathbf{s}$ than Eve. An overview of multi-dimensional reconciliation is given in Fig. \ref{Reconciliation}.

At the start of the CV-QKD protocol, Alice transmits a sequence $\mathbf{x} = [x^I_1, x^Q_1, \cdots, x^I_{N/2}, x^Q_{N/2}]$  of length $N$ over the quantum channel, where $I$ and $Q$ refer to the in-phase and quadrature components of the quantum states. Thus, $[x_i^I,x_i^Q]$ corresponds to a constellation point in a constellation $\mathcal{X}$ and is randomly sampled using a quantum random number generator (QRNG). For the rest of the paper, we will write the sequence as $\mathbf{x} = [x_1,x_2,\cdots,x_{N}]$ such that $[x_{2i-1}, x_{2i}] = [x_i^I,x_i^Q] \ \ \ \forall i \in 1,2,\cdots,N/2$. 
In the quantum channel, which is assumed to be an additive white Gaussian noise (AWGN) channel, noise $\mathbf{z}$ gets added to $\mathbf{x}$. This noise is Gaussian-distributed with a distribution of $\mathcal{N}(0,\sigma_z^2/2)$, where $\sigma_z^2$ is the total noise variance over both the in-phase and quadrature components of the noise. 

Using a coherent quantum receiver, Bob measures the quantum symbols and obtains a sequence $\mathbf{y} = \mathbf{x} + \mathbf{z}$. He generates a random bit string $\mathbf{s}$ using QRNG of length $N\cdot R$, where $R$ is the rate of the error correction code. Using an encoder, Bob encodes $\mathbf{s}$, creating the sequence $\mathbf{c}$, which is a codeword from the family of codewords $\mathcal{C}$ from the error correction code. He transforms the bits of his codewords to a sequence of BPSK symbols $\mathbf{u}$ such that $u_i = (-1)^{c_i} \ \ \ \forall i \in 1,2,\cdots,N$.
Bob uses a mapping function $M(\mathbf{u},\mathbf{y})$ which maps $\mathbf{u}$ and $\mathbf{y}$ to a sequence $\mathbf{m}$ of length $N$ such that applying the inverse of the function to $\mathbf{m}$ and $\mathbf{y}$ will give $\mathbf{u}$, i.e., $M^{-1}(\mathbf{m},\mathbf{y}) = \mathbf{u}$. More detail on the mapping function is given in \cite{Leverrier_2008}. Bob transmits $\mathbf{m}$ over the classical channel, which is assumed to be error-free and openly accessible to Eve, to Alice. 

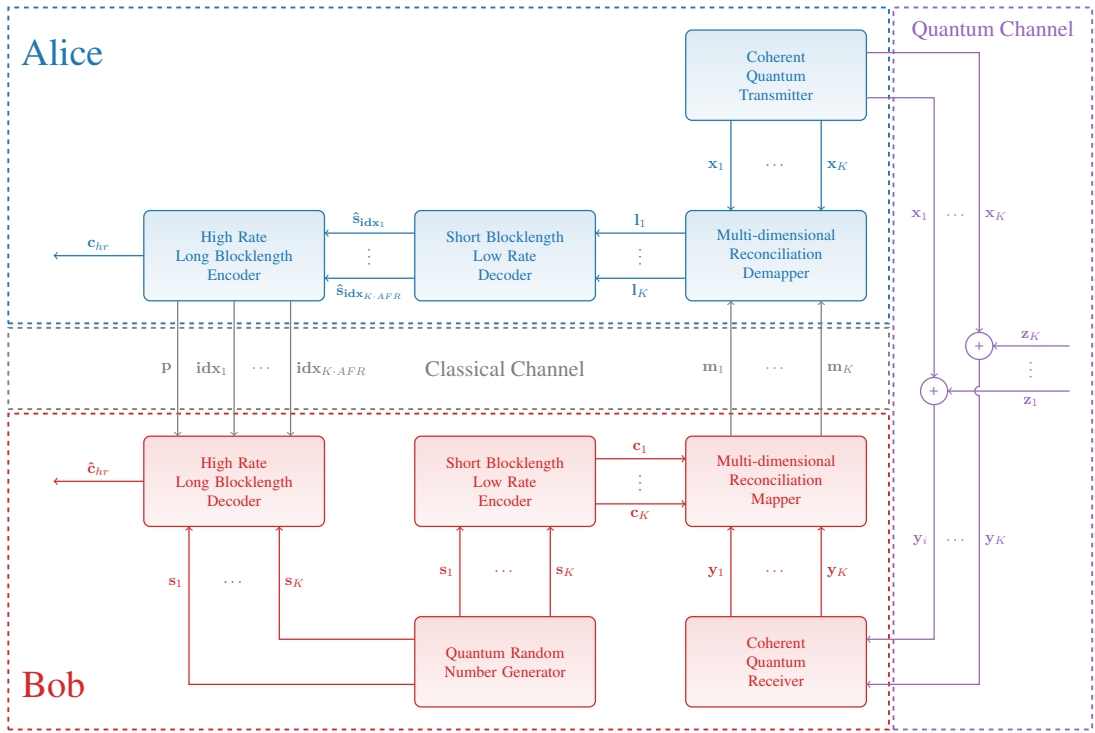
\begin{figure*}
    \centering
    \resizebox{0.8\linewidth}{!}{\begin{tikzpicture}
\draw[dashed, very thick, C6](14.6,2.5) rectangle (19,-13.5);
\node[C6] at (16.8,2){\Large Quantum Channel}; 
\draw[dashed, very thick, C8](-5,-4.6) rectangle (14.5,-6.4);
\node[C8] at (6,-5.5){\Large Classical Channel};
    \draw[dashed,very thick, C1](-5,-4.5) rectangle (14.5,2.5);

    \node[C1] at (-3.8,1.5){\Huge Alice};

    \draw[rounded corners,thick,C1,gradientc1] (10,0) rectangle node[midway, align = center]{Coherent\\ Quantum \\ Transmitter} (14,2);

    \draw[rounded corners,thick,C1,gradientc1] (10,-2) rectangle node[midway, align = center]{Multi-dimensional\\ Reconciliation \\ Demapper} (14,-4);

    \draw[rounded corners,thick,C1,gradientc1] (4,-2) rectangle node[midway, align = center]{Short Blocklength\\ Low Rate \\ Decoder} (8,-4);

    \draw[rounded corners,thick,C1,gradientc1] (-2,-2) rectangle node[midway, align = center]{High Rate\\ Long Blocklength \\ Decoder} (2,-4);
    \draw[dashed,very thick,C4](-5,-6.5) rectangle (14.5,-13.5);

    \node[C4] at (-4,-12.5){\Huge Bob};

    \draw[rounded corners,thick,C4,gradientc4] (10,-11) rectangle node[midway, align = center]{Coherent\\ Quantum \\ Receiver} (14,-13);

    \draw[rounded corners,thick,C4,gradientc4] (4,-9) rectangle node[midway, align = center]{Short Blocklength\\ Low Rate \\ Encoder} (8,-7);

    \draw[rounded corners,thick,C4,gradientc4] (4,-11) rectangle node[midway, align = center]{Quantum Random \\ Number Generator} (8,-13);
    
    \draw[rounded corners,thick,C4,gradientc4] (10,-7) rectangle node[midway, align = center]{Multi-dimensional\\ Reconciliation \\ Mapper} (14,-9);
        
    \draw[rounded corners,thick,C4,gradientc4] (2,-7) rectangle node[midway, align = center]{High Rate\\ Long Blocklength \\ Encoder} (-2,-9);


\draw[thick,->,C1](11,0) -- node[midway,left]{$\mathbf{x}_1$} (11,-2);

\node[C1] at (12,-1){$\cdots$};

\draw[thick,->,C1](13,0) -- node[midway,right]{$\mathbf{x}_K$} (13,-2);

\draw[thick,->,C6](14,0.5) -- (15.5,0.5) -- node[midway,left,yshift= 5mm]{$\mathbf{x}_1$} (15.5,-5.7);

\node[C6] at (16,-2.1){$\cdots$};

\draw[thick,->,C6](14,1.5) -- (16.5,1.5) -- node[midway,right,yshift = -5mm]{$\mathbf{x}_K$} (16.5,-4.7);

\draw[thick,C6] (15.5,-6) circle (0.3cm);
\node[C6] at (15.5,-6){+};

\draw[thick,C6] (16.5,-5) circle (0.3cm);
\node[C6] at (16.5,-5){+};

\draw[thick,->,C6] (18.5,-6) -- node[midway,below,xshift = 5mm]{$\mathbf{z}_1$}(15.8,-6);
\node[rotate = 90,C6] at (17.65,-5.5){$\cdots$};
\draw[thick,->,C6] (18.5,-5) -- node[midway,above]{$\mathbf{z}_K$}(16.8,-5);

\draw[thick,->,C6](15.5,-6.3) -- node[midway,left,yshift = -4mm]{$\mathbf{y}_i$}(15.5,-11.5) -- (14,-11.5);

\node[C6] at (16,-9.3){$\cdots$};

\draw[thick,->,C6](16.5,-5.3) -- (16.5,-5.9) arc(90:270:0.1) -- node[midway,right]{$\mathbf{y}_K$}(16.5,-12.5) -- (14,-12.5);

\draw[thick,->,C4](11,-11) -- node[midway,left]{$\mathbf{y}_1$} (11,-9);

\node[C4] at (12,-10){$\cdots$};

\draw[thick,->,C4](13,-11) -- node[midway,right]{$\mathbf{y}_K$} (13,-9);

\draw[thick,->,C8](11,-7) -- node[midway,left]{$\mathbf{m}_1$} (11,-4);

\node[C8] at (12,-5.5){$\cdots$};

\draw[thick,->,C8](13,-7) -- node[midway,right]{$\mathbf{m}_K$} (13,-4);

\draw[thick,->,C4](5,-11) -- node[midway,left]{$\mathbf{s}_1$} (5,-9);

\node[C4] at (6,-10){$\cdots$};

\draw[thick,->,C4](7,-11) -- node[midway,right]{$\mathbf{s}_K$} (7,-9);

\draw[thick,->,C4](8,-8.5) -- node[midway,below]{$\mathbf{c}_K$} (10,-8.5);

\node[rotate = 90,C4] at (9,-8){$\cdots$};

\draw[thick,->,C4](8,-7.5) -- node[midway,above]{$\mathbf{c}_1$} (10,-7.5);

\draw[thick,<-,C1](8,-2.5) -- node[midway,above]{$\mathbf{l}_1$} (10,-2.5);

\node[rotate = 90,C1] at (9,-3){$\cdots$};

\draw[thick,<-,C1](8,-3.5) -- node[midway,below]{$\mathbf{l}_K$} (10,-3.5);

\draw[thick,->,C1](4,-2.5) -- node[midway,above]{$\mathbf{\hat{s}}_{\mathbf{idx}_1}$} (2,-2.5);

\node[rotate = 90,C1] at (3,-3){$\cdots$};

\draw[thick,->,C1](4,-3.5) -- node[midway,below]{$\mathbf{\hat{s}}_{\mathbf{idx}_{A}}$} (2,-3.5);

\draw[thick,<-,C8](0,-7) -- node[midway,left]{$\mathbf{idx}_1$} (0,-4);

\node[C8] at (0.625,-5.5){$\cdots$};

\draw[thick,<-,C8](1.25,-7) -- node[midway,right]{$\mathbf{idx}_{A}$} (1.25,-4);

\draw[thick,->,C4](4,-11.5) -- (1,-11.5) -- node[midway,right]{$\mathbf{s}_K$} (1,-9);

\node[C4] at (0,-10.25){$\cdots$};

\draw[thick,->,C4](4,-12.5) -- (-1,-12.5) -- node[midway,left,yshift = 5mm]{$\mathbf{s}_1$} (-1,-9);

\draw[thick,<-,C8](-1.25,-4) -- node[midway,left]{$\mathbf{p}$} (-1.25,-7);

\draw[thick,->,C1] (-2,-3) -- node[midway,above]{$\mathbf{\hat{w}}$}(-4,-3);

\draw[thick,->,C4] (-2,-8) -- node[midway,above]{$\mathbf{w}$}(-4,-8);
\end{tikzpicture}}
    \caption{An overview of our proposed reconciliation protocol based on multi-dimensional reconciliation.}    
    \label{fig:Protocol}
\end{figure*} 

Alice receives $\mathbf{m}$ and applies the inverse of the mapping function using $\mathbf{x}$ to get $\mathbf{r} = M^{-1}(\mathbf{m},\mathbf{x})$. Because $\mathbf{y}$ is a noisy version of $\mathbf{x}$, the demapped result $\mathbf{r}$ will not be equal to $\mathbf{u}$. Instead, a virtual channel has been created where $\mathbf{r} = \mathbf{u} + \mathbf{n}$, where $\mathbf{n}$ is the noise of the virtual channel. To retrieve $\mathbf{c}$, error correction needs to be performed to get rid of the noise. Alice calculates the log-likelihood ratios (LLRs) $\mathbf{l}$ of her received message and uses these LLRs to attempt to decode the codeword. After decoding, she will be left with $\mathbf{\hat{c}}$, an estimate of $\mathbf{c}$. To check whether the codeword was decoded correctly, Alice first checks whether the syndrome of $\mathbf{\hat{c}}$ is equal to $\mathbf{0}$, i.e, $\mathbf{\hat{c}}\mathbf{H}^T = \mathbf{0}$, where $\mathbf{H}$ is the parity check matrix of the error correction code. If this is not the case, a frame error has occurred, and Alice discards the frame. If the syndrome is equal to $\mathbf{0}$, $\mathbf{\hat{c}}$ is a valid codeword of $\mathcal{C}$, however this does not guarantee that $\mathbf{\hat{c}} = \mathbf{c}$.

One final confirmation step is done by performing a universal hash function on $\mathbf{\hat{s}}$, the information bits of $\mathbf{\hat{c}}$, and transmitting the result $h_{\mathbf{\hat{s}}}$ to Bob. 
Bob compares $h_{\mathbf{\hat{s}}}$ to the hashing result of $\mathbf{s}$, $h_{\mathbf{s}}$. 
If they are the same, Alice and Bob can say with very high confidence that $\mathbf{\hat{s}} = \mathbf{s}$, and they will use these bit strings to distil keys during privacy amplification. If the hashing results are not the same, a frame error has occurred, and the entire frame is discarded. This hashing reveals some information on the bits, as $h_{\mathbf{\hat{s}}}$ is transmitted over the classical channel. 
This reduces the total SKR, but, because the blocklengths of the error correction codes are quite long, this leakage of information is negligible. The amount of information leaked depends on the hashing method; for example, in \cite{Milicevic_2018}, a 32-bit cyclic redundancy check (CRC) is used to hash an $R = \frac{1}{50}$ code with $N = 10^6$. The CRC bits are discarded after hashing, so they are not used for key distillation. Hence, the total rate of the code, and thus the reconciliation efficiency, decreases slightly and becomes $R' = \frac{RN-32}{N} = 0.019968$, a 0.16\% decrease in rate. This decreases the reconciliation efficiency, defined as $\beta = \frac{R'}{I_{AB}}$, where $I_{AB}$ is the mutual information between Alice and Bob, by approximately 0.16\% as well, which has some impact on the SKR, but is mostly negligible. When the blocklength is smaller, however, discarding the revealed bits has a significant influence on $\beta$, e.g., for $N= 10^4$, $R = \frac{1}{50}$ and 32 CRC bits, the rate decreases by 16\%. 


The following equation has been suggested for the SKR for the multi-dimensional reconciliation protocol \cite{Milicevic_2018}:
\begin{equation}
    \text{SKR} = (1-\text{FER})(\beta I_{AB} - \chi_{BE} - \Delta_N),
    \label{eq:SKR}
\end{equation}
where $\chi_{BE}$ is the Holevo information between Eve and Bob, and $\Delta_N$ is the finite-size penalty of the privacy amplification, which depends on the privacy amplification blocklength $N_{privacy}$
\begin{equation}
    \Delta_N = 7\sqrt{\frac{\log_2(2/\epsilon)}{N_{privacy}}},
\end{equation}
where $\epsilon$ is the security parameter, which is often chosen to be $10^{-10}$ \cite{leverrier2010finite}. 

\section{Proposed Reconciliation Protocol}
\label{Section:Proposed Reconciliation}
Our proposed protocol is based on both the multi-dimensional reconciliation and the reconciliation protocol using random codebooks in \cite{ray2025random} and uses short blocklength error correction codes. Although this protocol can be used for $\beta \leq 1$, in this paper we focus on $\beta > 1$. 

The implication of operating with $\beta > 1$ is that, although the total information throughput decreases because of the high FER, the total secret information that is shared increases because Alice is capable of extracting relatively more information from the accepted frames compared to Eve, who can only ever extract $\chi_{BE}$ per bit from the accepted frames. By operating at a very high FER, we are only accepting frames that the decoder can decode. These decoded frames will not necessarily be decoded correctly, but will, on average, have a significantly lower error rate than the rejected frames. The residual errors in the accepted frames can then be corrected using a high-rate code. This principle is similar to the advantage distillation used in classical cryptography \cite{Maurer1993} and device-independent QKD protocols \cite{Tan2020}. 

As shown in \cite{johnson2017problem}, it is possible to operate with $\beta > 1$ using the standard long blocklength LDPC codes with a $\beta = 1.09$ with an FER of 0.9999. However, this long blocklength significantly reduces the performance in this regime. Normally, long blocklengths are used as they approach the performance of infinite blocklength codes \cite{shannon1948mathematical}. However, operating in the finite blocklength regime is preferable from an FER perspective when the code rate is above capacity, as the strong converse of the coding theorem states that the FER approaches 1 as the blocklength tends to infinity \cite{feinstein1959coding}. Therefore, a small blocklength is desirable when $\beta > 1$ to reduce the FER as much as possible. One downside, however, is that for short blocklength codewords, it is not possible to confirm that a codeword was decoded correctly using a universal hash function without significantly reducing the code rate, as discussed before. There are potential security implications of operating with $\beta > 1$, as also noted in \cite{johnson2017problem}, which we will discuss in a later section. 

In Fig. \ref{fig:Protocol}, an overview of our proposed protocol is shown. The protocol has two decoding steps, the first with a short blocklength low-rate inner code, the second with a long blocklength high-rate outer code. The blocklength of the outer high rate code $N_{out}$ is chosen as a multiple of the information length of the inner code $N_{in}$, $N_{out} = A\cdot N_{in}\cdot R_{in}$, where $R_{in}$ is the rate of the short blocklength code. At the start of the protocol, Alice and Bob perform multi-dimensional reconciliation with $N_{in}$ chosen to be very small, and $\beta_{in}$, the reconciliation efficiency of the first decoding step, $ > 1$. Instead of confirming whether the codeword is correctly decoded using the syndrome of the LDPC code and the CRC bits, Alice orders these codewords based on the log a-posteriori probability ratios at the output of the decoder $\mathbf{l}_{i,out}$. For each codeword, Alice calculates $q_i = \sum_{j = 1}^{N_l}{|l_{i,out_j}|}$. The higher $q_i$ is, the more certain the decoder is about the correctness of the decoded codeword. This is because the log a-posteriori probabilities increase in magnitude as the decoder converges. After sorting the codewords based on $q_i$, Alice decides to accept only a fraction of the codewords with the highest $q_i$. This fraction is the accepted frame rate (AFR) and equals $(1-\text{FER})$ in standard reconciliation. The cut-off value $q_c$ for which to accept or reject decoded codewords to obtain a given AFR can be determined through simulations. Then, all decoded codewords for which $q \leq q_c$ are accepted, while the others are discarded. It is important to note that the decoding of a codeword does not depend on any other codeword.
The number of reconciliation attempts $K$ required to have enough bits for the second error correction step is a random variable with $\mathbb{E}[K] = \frac{N_{out}}{N_{in} \cdot R_{in} \cdot \text{AFR}}$. 

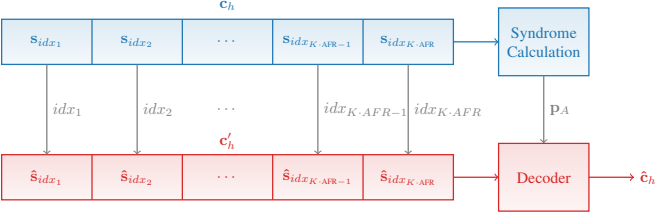
\begin{figure}[b!]
    \centering
    \resizebox{\linewidth}{!}{\begin{tikzpicture}

    \draw[thick,C4,gradientc4] (-1,-1.5) rectangle (9,-2.5);   
    \foreach \x in {1,3,5,7}{
    \draw[thick,C4] (\x,-1.5) -- (\x,-2.5);}

    \node[C4] at (0,-2) {$\mathbf{s}_{idx_1}$};
    \node[C4] at (2,-2) {$\mathbf{s}_{idx_2}$};
    \node[C4] at (4,-2) {$\cdots$};
    \node[C4] at (6,-2) {$\mathbf{s}_{idx_{A-1}}$};
    \node[C4] at (8,-2) {$\mathbf{s}_{idx_{A}}$};

    \node[C4] at (4,-1.2) {$\mathbf{w}$};

    \draw[rectangle,thick,C4,gradientc4] (10,-2.75) rectangle (12,-1.25) node[midway, align = center]{Syndrome\\ Calculation};

    \draw[->,thick,C4](9,-2) -- (10,-2);

    \draw[->,thick,C8](11,-1.25) -- (11,0.25) node[midway,right]{$\mathbf{p}$};

    \draw[thick,C1,gradientc1] (-1,1.5) rectangle (9,0.5);   
    \foreach \x in {1,3,5,7}{
    \draw[thick,C1] (\x,0.5) -- (\x,1.5);}

    \node[C1] at (0,1) {$\mathbf{\hat{s}}_{idx_1}$};
    \node[C1] at (2,1) {$\mathbf{\hat{s}}_{idx_2}$};
    \node[C1] at (4,1) {$\cdots$};
    \node[C1] at (6,1) {$\mathbf{\hat{s}}_{idx_{A-1}}$};
    \node[C1] at (8,1) {$\mathbf{\hat{s}}_{idx_{A}}$};

    \node[C1] at (4,1.8) {$\mathbf{w}'$};

    \draw[rectangle,thick,C1,gradientc1] (10,0.25) rectangle (12,1.75) node[midway]{Decoder};

    \draw[->,thick,C1](9,1) -- (10,1);

    \draw[->,thick,C1](12,1) -- (13,1) node[right]{$\mathbf{\hat{w}}$};

    \draw[->,thick,C8](0,0.5) -- (0,-1.5) node[midway,right]{$idx_1$};
    \draw[->,thick,C8](2,0.5) -- (2,-1.5) node[midway,right]{$idx_2$};
    \draw[->,thick,C8](6,0.5) -- (6,-1.5) node[midway,right]{$idx_{A-1}$};
    \draw[->,thick,C8](8,0.5) -- (8,-1.5) node[midway,right]{$idx_{A}$};
    \node[C8] at (4,-0.5){$\cdots$};
    \end{tikzpicture}}
    \caption{An overview of the second error correction step using a high-rate code.}    
    \label{fig:Encoding}
\end{figure}

The accepted frames are not necessarily decoded correctly, but they have a relatively low bit-error rate (BER). Normally, a hashing function is applied to both Alice's and Bob's information bits, but as mentioned before in the short blocklength case, this would significantly impact the reconciliation efficiency. Therefore, we need to correct the residual bit errors that remain in the accepted information bits. To achieve this, Alice concatenates all of the estimated information bits together to create one very long string of bits $\mathbf{w}'$ of length $N_{out}$, as shown in Fig. \ref{fig:Encoding}. Alice transmits the indices of all accepted codewords $\mathbf{idx}$ to Bob. Bob concatenates the information bits of the accepted codewords to create $\mathbf{w}$, and calculates the syndrome $\mathbf{p}$ of this bit sequence using the parity check matrix $\mathbf{H}_{out}$ of the high-rate code, and transmits a one-time padded version of this syndrome to Alice. This syndrome needs to be calculated and transmitted to Alice, as $\mathbf{w}$ is (with high probability) not a valid codeword of the high-rate code.

Although $\mathbf{w}$ is not a valid codeword of the high-rate code, in combination with the syndrome $\mathbf{p}$, Alice can decode to $\mathbf{\hat{w}}$ as we will show in the following. 
 Let $\mathbf{t}$ be any arbitrary codeword from the family of codewords $\mathcal{C}_{out}$ from the high-rate error correction code. In that case, Alice's bit string $\mathbf{w}'$ is equal to $\mathbf{t} \oplus \mathbf{a}$, where $\mathbf{a}$ is a binary sequence indicating the bit positions at which $\mathbf{w}'$ and $\mathbf{t}$ are different. Similarly, Bob's bit string is $\mathbf{w}= \mathbf{t} \oplus \mathbf{b}$. As mentioned before, $\mathbf{w}'$ is equal to $\mathbf{w}$ with some bit flips, i.e., $\mathbf{w}'= \mathbf{w} \oplus \mathbf{e}$, where $\mathbf{e}$ indicates the positions of the erroneous bits. Therefore, $\mathbf{w} = \mathbf{t} \oplus \mathbf{a} \oplus \mathbf{e}$. When the syndromes of $\mathbf{w}$ ($\mathbf{p}$) and $\mathbf{w}'$ ($\mathbf{p}'$) are added together, the result will be the syndrome of the error pattern $\mathbf{e}$.
The complete mathematical derivation is given below:
\begin{align}
    \mathbf{w}' &= \mathbf{t} \oplus \mathbf{a}\\\nonumber
    \mathbf{w} &= \mathbf{t} \oplus \mathbf{b} = \mathbf{t} \oplus \mathbf{a} \oplus \mathbf{e}\\\nonumber
    \mathbf{p} &= \mathbf{w}\mathbf{H}_{out}^T\\\nonumber
    \mathbf{p}' &=  \mathbf{w}'\mathbf{H}_{out}^T\\\nonumber
    \mathbf{p} \oplus \mathbf{p}' &= \mathbf{w}\mathbf{H}_{out}^T \oplus \mathbf{w}'\mathbf{H}_{out}^T \\\nonumber
    &= (\mathbf{t} \oplus \mathbf{a} \oplus \mathbf{t} \oplus \mathbf{a} \oplus \mathbf{e})\mathbf{H}_{out}^T\\\nonumber
    &= \mathbf{e}\mathbf{H}_{out}^T.
\end{align}
Therefore, the binary addition of $\mathbf{p}$ and $\mathbf{p}'$ can be used in the decoder to get $\mathbf{\hat{e}}$, which is an estimate of $\mathbf{e}$. She applies it to $\mathbf{w}'$ to get $\mathbf{\hat{w}} = \mathbf{w}' \oplus \mathbf{\hat{e}}$, which is an estimate of $\mathbf{w}$.

The FER of the second step is chosen to be very low (FER < $10^{-9}$), such that Alice and Bob can be sure, with a probability close to 1, that their bit strings are the same. Therefore, an additional hashing step to confirm the correctness of the decoding is not necessary, but could be performed optionally. 

For our proposed protocol, we need to use secret key material when performing one-time padding of $\mathbf{p}$ for transmission over the classical channel. In the following, we will show that this usage of key material is equivalent to a reduction in reconciliation efficiency.

The length of a syndrome for any arbitrary block code is equal to $N(1-R)$. For the high rate code, the blocklength is $N_{out} = K \text{AFR} N_{in}  \beta_{in}  I_{AB}$, where $\beta_{in} = \frac{R_{in}}{I_{AB}}$. The rate of the code depends on the amount of bit errors in the accepted frames $\text{BER}_{AF}$, which can statistically be determined by performing simulations for a given channel, code, and AFR. The capacity of the binary symmetric channel (BSC) created by discarding and concatenating the low-rate codewords is $1-h(\text{BER}_{AF})$, where $h(x)$ is the binary entropy function. The rate of the high rate code is determined to be $R_{out} = \beta_{out} (1-h(\text{BER}_{AF}))$, where $\beta_{out} = \frac{R_{out}}{(1-h(\text{BER}_{AF}))}$ is the reconciliation efficiency of the high rate code. Therefore, the length of $\mathbf{p}$ is equal to $N_{out}(1-R_{out}) = K \text{AFR} N_{in}  \beta_{in}  I_{AB} \beta_{out} h(\text{BER}_{AF})$, which if we normalise it to the amount of key material used per bit transmitted over the classical channel, where we transmit a total of $KN_{in}$ bits, becomes $\text{AFR}\beta_{in}  I_{AB} \beta_{out} h(\text{BER}_{AF})$.
 The secret key rate for our proposed protocol is then:
\begin{align}
         \text{SKR}_{t} &= (1-\text{FER}_{out})(\text{AFR}(\beta_{in}I_{AB} - \chi_{BE}) \\ &-\text{AFR}\beta_{in}  I_{AB} \beta_{out} h(\text{BER}_{AF}))  \nonumber\\
        \text{SKR}_{t} &= \text{AFR}(1-\text{FER}_{out})(\beta_
        {in}\beta_{out}(1-h(\text{BER}_{AF}))I_{AB}-\chi_{BE}) \nonumber\\\nonumber
        \text{SKR}_{t} &= (1-\text{FER}_t)(\beta_tI_{AB}-\chi_{B}),
\end{align}
where $\text{FER}_t = (1-\text{AFR}(1-\text{FER}_{out}))$ is the total FER of the protocol, and $\beta_t = \beta_{in}\beta_{out}(1-h(\text{BER}_{AF}))$ is the total reconciliation efficiency of the protocol. Because the high rate code operates at a very low FER ($\text{FER}_{out} < 10^{-9}$), $(1-\text{FER}_t) \approx \text{AFR}$. It is important to note that this equation would only be valid if the security proofs from \cite{Leverrier_2008} still hold under these conditions, which we will discuss later.
\section{Results}
\label{Section: Results}
\begin{figure}[!b]
    \centering
      \begin{tikzpicture}
\begin{axis}[
every axis/.append style={font=\small},
tick label style={font=\footnotesize},
xlabel = AFR,
ylabel = BER$_{AF}$(---),
xmin = 0, xmax =0.1,
ymin = 0.0001, ymax = 1,
ymode = log,
x tick label style={yshift= -1mm},
y tick label style={xshift= -1mm},
ylabel shift = 1mm,
xlabel shift = 1mm,
width=0.88\linewidth,
height=7cm,
ytick pos = left,
set layers, mark layer=axis tick labels,
 xlabel near ticks,  
 ylabel near ticks,  
 xticklabel style={/pgf/number format/fixed},
every axis plot/.append style={thick},legend style={at={(0.35,0.2)},anchor=west, font = \scriptsize,row sep=-0.75ex,inner sep=0.2ex},
legend cell align={left},
cycle list name = foo
]

\pgfplotstableread{Figures/AFR_BER.txt}
\datatable
\foreach \x in {1,2,3,4,5,6}{
    \addplot+[no marks]
             table
             [
              x expr=\thisrowno{0}, 
              y expr=\thisrowno{\x} 
             ] {\datatable};
}
\addlegendentry{$\beta_{in} = 1.6$}
\addlegendentry{$\beta_{in} = 1.5$}
\addlegendentry{$\beta_{in} = 1.4$}
\addlegendentry{$\beta_{in} = 1.3$}
\addlegendentry{$\beta_{in} = 1.2$}
\addlegendentry{$\beta_{in} = 1.1$}
\end{axis}

\begin{axis}[
every axis/.append style={font=\small},
tick label style={font=\footnotesize},
xlabel = AFR,
ylabel = $1-h(\text{BER}_{AF})$ (\--\--\--),
xmin = 0, xmax =0.1,
ymin = 0, ymax = 1,
axis y line*=right,
x tick label style={yshift= -1mm},
y tick label style={xshift= -1mm},
ylabel shift = 1mm,
xlabel shift = 1mm,
width=0.88\linewidth,
axis x line=none,
height=7cm,
set layers, mark layer=axis tick labels,
 xlabel near ticks,  
 ylabel near ticks,  
 xticklabel style={/pgf/number format/fixed},
every axis plot/.append style={thick},legend style={at={(0.45,0.2)},anchor=west, font = \scriptsize,row sep=-0.75ex,inner sep=0.2ex},
legend cell align={left},
cycle list name = foo
]

\pgfplotstableread{Figures/AFR_R.txt}
\datatable
\foreach \x in {1,2,3,4,5,6}{
    \addplot+[no marks, dashed]
             table
             [
              x expr=\thisrowno{0}, 
              y expr=\thisrowno{\x} 
             ] {\datatable};
}
\end{axis}
\end{tikzpicture}
    \caption{$\text{BER}_{AF}$ (---) and $1-h(\text{BER}_{AF})$ (\--\--\--) vs. AFR for different $\beta_{in}$. The error correction code used is a $R_{in} = \frac{1}{50}$ TBP-LDPC code with $N_{in} = 500$.}    
    \label{fig:AFR_BER}
\end{figure}
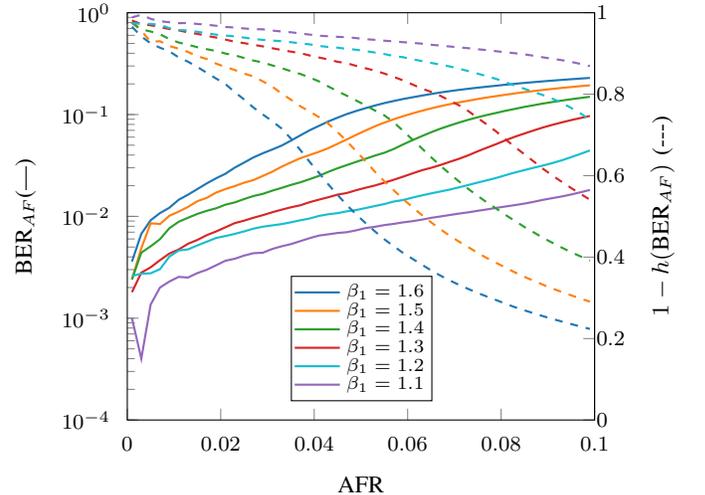
To show the performance of the proposed protocol, we have simulated the protocol assuming the use of short blocklength LDPC codes for the first decoding step. For the second decoding step, we take into account two different scenarios: one where the second decoding step is perfect ($\beta_{out} = 1$,$\text{FER}_{out} = 0$) and one where it is sub-optimal ($\beta_{out} = 0.9$, $\text{FER}_{out} = 0$).

\begin{figure}[!t]
    \centering
      \begin{tikzpicture}
\begin{axis}[
every axis/.append style={font=\small},
tick label style={font=\footnotesize},
xlabel = AFR,
ylabel = $\beta_{t}$,
xmin = 0, xmax =0.1,
ymin = 1, ymax = 1.6,
x tick label style={yshift= -1mm},
y tick label style={xshift= -1mm},
ylabel shift = 1mm,
xlabel shift = 1mm,
width=\linewidth,
height=7cm,
ytick pos = left,
set layers, mark layer=axis tick labels,
 xlabel near ticks,  
 ylabel near ticks,  
 grid = major,
 xticklabel style={/pgf/number format/fixed},
every axis plot/.append style={thick},legend style={at={(0.6,0.5)},anchor=west, font = \scriptsize,row sep=-0.75ex,inner sep=0.2ex},
legend cell align={left},
cycle list name = foo
]

\pgfplotstableread{Figures/AFR_Beta_total.txt}
\datatable
\foreach \x in {1,2,3,4,5,6}{
    \addplot+[no marks]
             table
             [
              x expr=\thisrowno{0}, 
              y expr=\thisrowno{\x} 
             ] {\datatable};
} 
\pgfplotstableread{Figures/AFR_Beta.txt}
\datatable
    \addplot[no marks, black,dashed]
             table
             [
              x expr=\thisrowno{0}, 
              y expr=\thisrowno{1} 
             ] {\datatable};
\addlegendentry{$\beta_l = 1.6$}
\addlegendentry{$\beta_l = 1.5$}
\addlegendentry{$\beta_l = 1.4$}
\addlegendentry{$\beta_l= 1.3$}
\addlegendentry{$\beta_l = 1.2$}
\addlegendentry{$\beta_l = 1.1$}
\addlegendentry{Optimal}
\end{axis}

\end{tikzpicture}
    \caption{$\beta_t$ vs. AFR for our proposed protocol for different $\beta_{in}$, assuming $\beta_{out} = 1$. The error correction code used is a $R_{in} = \frac{1}{50}$ TBP-LDPC code with $N_{in} = 500$.}      
    \label{fig:AFR_Beta_total}
\end{figure}
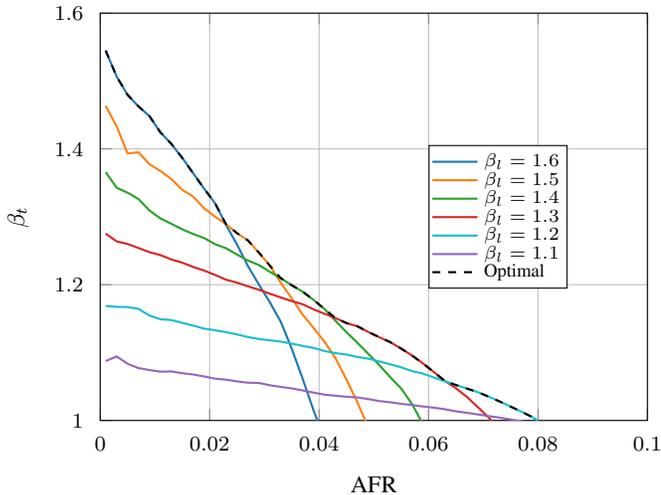

In Fig. \ref{fig:AFR_BER} we show the AFR against both BER$_{AF}$ and $1-h(\text{BER}_{AF})$, assuming the use of a type-based protograph (TBP) LDPC code with $R_{in}= \frac{1}{50}$ with $N_{in} = 500$ taken from \cite{gumucs2021low}. The parity check matrices in this paper were not optimised for short cycles, so the performance reported here is not as good as it could be. Because of these short cycles, certain codewords with many bit errors have high $q_i$ and are therefore accepted. This can be prevented by optimising the parity-check matrix or by initially applying a parity check to determine whether the codeword is valid, as in \cite{gumucs2025short}; however, that is outside the scope of this paper. Simulation results are shown for different values of $\beta_{in}$. As shown, when the AFR decreases, BER$_{AF}$ decreases as well. This is because we order the frames based on the $\mathbf{l}_{i,out}$, i.e., based on how certain the decoder is that a particular codeword was decoded. The fewer frames we accept, the more certain we are about the accepted codewords, hence a lower BER is achieved. Conversely, the channel capacity of the resulting BSC increases as the AFR decreases. When $\beta_{in}$ increases, BER$_{AF}$ increases as well, meaning that a lower rate correction code is required during the second decoding step. 

\begin{figure}[!b]
    \centering
      \begin{tikzpicture}
\begin{axis}[
every axis/.append style={font=\small},
tick label style={font=\footnotesize},
ylabel = $\beta_t$,
xlabel = AFR,
xmin = 0, xmax =0.1,
ymin = 1, ymax = 1.6,
x tick label style={yshift= -1mm},
y tick label style={xshift= -1mm},
ylabel shift = 1mm,
xlabel shift = 1mm,
width=\linewidth,
height=7cm,
ytick pos = left,
set layers, mark layer=axis tick labels,
 xlabel near ticks,  
 ylabel near ticks,  
 xticklabel style={/pgf/number format/fixed},
every axis plot/.append style={thick},legend style={at={(0.3,0.8)},anchor=west, font = \scriptsize,row sep=-0.75ex,inner sep=0.2ex},
legend cell align={left},
cycle list name = foo
]
\addplot+[only marks] coordinates{(0.0001,1.09)};
\pgfplotstableread{Figures/AFR_Beta.txt}
\datatable
    \addplot+[no marks]
             table
             [
              x expr=\thisrowno{0}, 
              y expr=\thisrowno{1} 
             ] {\datatable};
    \addplot+[no marks]
             table
             [
              x expr=\thisrowno{0}, 
              y expr=\thisrowno{2} 
             ] {\datatable};
\addlegendentry{\cite{johnson2017problem} $\beta = 1.09$, FER $= 0.9999$}
\addlegendentry{Our method $\beta_{out} = 1$, FER$_{out}$ = 0}
\addlegendentry{Our method $\beta_{out} = 0.9$, FER$_{out}$ = 0}
\end{axis}
\end{tikzpicture}
    \caption{$\beta_t$ vs. AFR for our proposed protocol with optimised $\beta_{in}$ compared to the state-of-the-art. The error correction code used is a $R_{in} = \frac{1}{50}$ TBP-LDPC code with $N_{in} = 500$.}    
    \label{fig:AFR_Beta}
\end{figure}
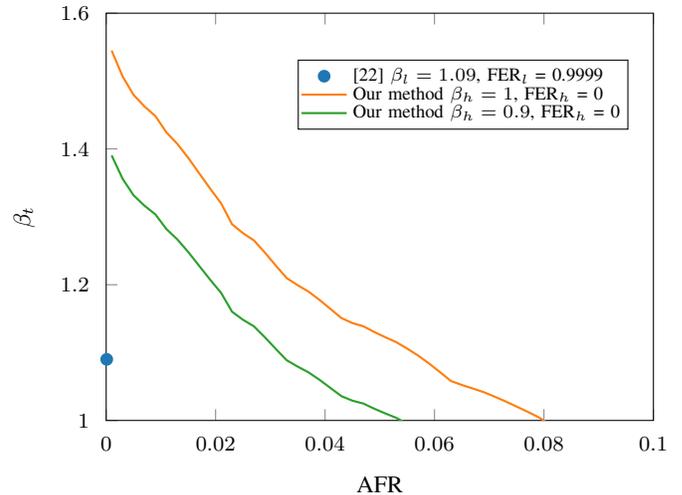

In Fig. \ref{fig:AFR_Beta_total}, we show that the optimal $\beta_{in}$ to choose depends on the target AFR. As we increase AFR, the capacity of the BSC increases faster for the higher $\beta_{in}$ than for the lower ones. As a result, at some point, the $R_{out}$ will be so low that choosing a lower $\beta_{in}$ will lead to a higher $\beta_t$ for the same AFR. In general, though, we want $\beta_t$ to be as high as possible, and we do not care as much about the AFR as we want to increase the distance of the CV-QKD system. 

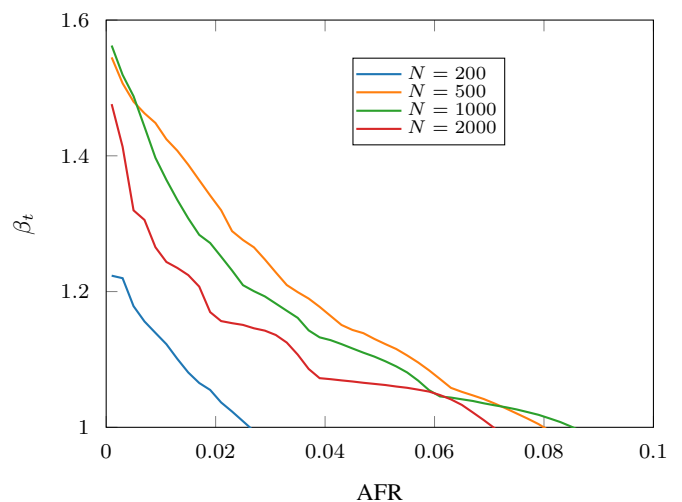
\begin{figure}[!t]
    \centering
      \begin{tikzpicture}
\begin{axis}[
every axis/.append style={font=\small},
tick label style={font=\footnotesize},
ylabel = $\beta_t$,
xlabel = AFR,
xmin = 0, xmax =0.1,
ymin = 1, ymax = 1.6,
x tick label style={yshift= -1mm},
y tick label style={xshift= -1mm},
ylabel shift = 1mm,
xlabel shift = 1mm,
width=\linewidth,
height=7cm,
ytick pos = left,
set layers, mark layer=axis tick labels,
 xlabel near ticks,  
 ylabel near ticks,  
 xticklabel style={/pgf/number format/fixed},
every axis plot/.append style={thick},legend style={at={(0.45,0.8)},anchor=west, font = \scriptsize,row sep=-0.75ex,inner sep=0.2ex},
legend cell align={left},
cycle list name = foo
]
\pgfplotstableread{Figures/AFR_Beta_Blocklength.txt}
\datatable
    \addplot+[no marks]
             table
             [
              x expr=\thisrowno{0}, 
              y expr=\thisrowno{1} 
             ] {\datatable};
     \addplot+[no marks]
             table
             [
              x expr=\thisrowno{0}, 
              y expr=\thisrowno{2} 
             ] {\datatable};
    \addplot+[no marks]
             table
             [
              x expr=\thisrowno{0}, 
              y expr=\thisrowno{3} 
             ] {\datatable};
    \addplot+[no marks]
             table
             [
              x expr=\thisrowno{0}, 
              y expr=\thisrowno{4} 
             ] {\datatable};
\addlegendentry{$N = 200$}
\addlegendentry{$N = 500$}
\addlegendentry{$N = 1000$}
\addlegendentry{$N = 2000$}
\end{axis}
\end{tikzpicture}
    \caption{$\beta_t$ vs. AFR for our proposed protocol for different $N_{in}$ assuming $\beta_{out} = 1$.The error correction code used is a $R_{in} = \frac{1}{50}$ TBP-LDPC code. }    
    \label{fig:AFR_Beta_Blocklength}
\end{figure}

In Fig. \ref{fig:AFR_Beta}, we show $\beta_t$ as a function of the AFR, optimised over our possible choices of $\beta_{in}$, comparing ideal decoding in the second step with sub-optimal decoding. We also compare it to the results from \cite{johnson2017problem}. Even assuming very sub-optimal decoding, $\beta_t$ of up to 1.4 can still be achieved. When we compare it to standard reconciliation protocols, only a $\beta_t$ of up to 1.09 can be achieved with a very low AFR of 0.0001, which significantly throttles the achievable SKRs.    

We have also investigated how the blocklength of short-blocklength LDPC codes influences the performance of our protocol. These results are shown in Fig. \ref{fig:AFR_Beta_Blocklength}. As the blocklength decreases, the performance of the protocol increases as well, as was expected. However, when $N_{in}$ is too small, performance degrades significantly, as seen for $N_{in} = 200$. This is a consequence of the error correction code used, as the code from \cite{gumucs2021low} was designed for very large blocklengths. When $N_{in}$ becomes too small, the code's performance breaks down because the parity-check matrix becomes too dense due to variable nodes with very high degrees. Additionally, because of the smaller blocklength, if a frame was wrongly decoded, the relative amount of errors is much higher, e.g., if there is 1 bit error in an accepted frame the BER of that one frame is 0.25 when $N_{in} = 200$, while for $N_{in} = 500$ the BER would be 0.1. 

\section{Security Considerations}
\label{Section:Security concerns}
Although the results for $\beta$ vs. FER are quite promising, there are security issues that need to be addressed before any claims can be made about SKR or achievable distances. The security proofs for multi-dimensional reconciliation \cite{Leverrier_2008} do not specifically mention whether the blocklength of the error correction codes is allowed to be arbitrarily chosen. From the paper, one must assume that, for the finite-size effects penalty, the block length of the error-correcting codes must be the same as for privacy amplification and parameter estimation. Additionally, the rejection of frames is never mentioned, as it is most likely viewed as a form of post-selection. However, in works that consider error correction, such as \cite{Milicevic_2018,gumucs2023adaptive,wang2017efficient}, the error-correction blocklength is always treated separately from the privacy-amplification blocklength, and the FER is allowed to be arbitrarily high. Therefore, the question is whether it is valid to assume that we are allowed to have a different blocklength for the privacy amplification and the error correction, and whether we are allowed to reject frames.

The $I_{AB}$ and $\chi_{BE}$ that were determined during the parameter estimation give an indication of the expected amount of noise on the measured quantum states for both Bob and Eve, respectively. During reconciliation, we split the parameter estimation block into smaller frames for the error correction. Although the noise follows the same distribution for each frame, the total noise of each of these frames is slightly different, as the noise is stochastic. 
Bob's and Eve's measurements are correlated, so if there is less noise on Bob's measurements, there is less noise on Eve's measurements; hence, the Holevo information $\chi_{BE}$ is higher. So the amount of information Eve can extract per quantum-state measurement differs across frames due to fluctuations in the total amount of noise.

In the case of no frame rejection (FER = 0), this is not an issue. If we accept every single frame, we use every measured quantum state to generate the keys. Hence, $\chi_{BE}$ will average out to what was estimated during the parameter estimation when calculating the SKR. So when FER = 0, we are allowed to use any blocklength and eq. \ref{eq:SKR} for determining the SKR. However, when we introduce frame errors, we can not make the same assumptions anymore. The frames that are accepted have, on average, a lower total amount of noise, while rejected frames have a higher total amount of noise. Equivalently, Eve has more information on measurements from accepted states than from rejected states, hence $\chi_{BE}$ is higher for the accepted frames. 
Thus, whether a frame is accepted or rejected affects the magnitude of the Holevo information $\chi_{BE}$. This effect is difficult to estimate, and hence protocols with post-selection of quantum states are normally not used, as security cannot be guaranteed. Furthermore, post-selection for the finite-size case has not yet been proven secure against Gaussian attacks \cite{leverrier2010finite}. In \cite{hajomer2025experimental,johnson2017problem, pirandola2024improved}, leakages caused by the post-selection done by rejecting frames are considered in the SKR equation; however, further research is required to see how rejecting frames influences the Gaussianity of the quantum states for different error correction protocols.

A potential solution is to one-time pad the accept/reject message so that Eve does not know which frames are accepted. However, from our simulation results, we observe that the SKR of our system eventually stops decreasing with distance, even when we one-time pad the accept/reject messages. This would mean that for arbitrarily long distance CV-QKD links, the SKR of our system would be infinitely above the Pirandola-Laurenza-Ottaviani-Banchi bound\cite{pirandola2017fundamental}, which cannot be the case. Therefore, there is some additional leakage which we have not yet taken into account. This leakage is either related to fluctuations in noise caused by finite-size effects, as mentioned before, or has not yet been discovered. This shows that it is imperative to perform a proper security analysis of this protocol before any claims about SKRs can be made. Additionally, if the blocklength does indeed cause additional finite-size effects, and if frame rejection is counted as a form of post-selection, this would have severe implications on previous error correction papers, and would significantly worsen the error correction performance, pushing CV-QKD back in terms of achievable distance and SKRs. 

\section{Conclusion}
In this paper, we have proposed a reconciliation protocol using short-block-length error-correcting codes with two decoding steps. We have shown that using this protocol, it is possible to achieve reconciliation efficiencies above 1 even with high frame error rates, potentially allowing us to increase the secret key rate and distance of CV-QKD protocols. We also discuss security considerations for this protocol and reconciliation protocols in general, and conclude that there are issues regarding the security of rejecting frames which need to be resolved before any claims can be made regarding SKR performance.
\label{Section:Conclusion}

\printbibliography[notcategory=ignore]

\end{document}